Motor cortex causally contributes to

auditory word recognition following sensorimotor-enriched vocabulary training

*[†]Brian Mathias[a,b], [†]Andrea Klingebiel[b], Gesa Hartwigsen[c],

Leona Sureth[b], Manuela Macedonia[b,d], Katja M. Mayer[e], & Katharina von Kriegstein[a,b]

[a]Chair of Cognitive and Clinical Neuroscience, Faculty of Psychology, Technical University Dresden, Dresden, Germany

[b]Research Group Neural Mechanisms of Human Communication, Max Planck Institute for Human Cognitive and Brain Sciences, Leipzig, Germany

[c]Lise Meitner Research Group Cognition and Plasticity, Max Planck Institute for Human Cognitive and Brain Sciences, Leipzig, Germany

[d]Linz Center of Mechatronics, Johannes Kepler University Linz, Linz, Austria

[e]Department of Psychology, University of Münster, Münster, Germany

*Corresponding author

[†]Joint first authors

Corresponding author information:

Brian Mathias

Technical University Dresden

Chair of Cognitive and Clinical Neuroscience

Faculty of Psychology

Bamberger Str. 7

01187 Dresden

Germany

Email: brian.mathias@tu-dresden.de

**Abstract**

The role of the motor cortex in perceptual and cognitive functions is highly controversial. Here, we investigated the hypothesis that the motor cortex can be instrumental for translating foreign language vocabulary. Participants were trained on foreign language (L2) words and their native language translations over four consecutive days. L2 words were accompanied by complementary gestures (sensorimotor enrichment) or pictures (sensory enrichment). Following training, participants translated the auditorily-presented L2 words that they had learned and repetitive transcranial magnetic stimulation (rTMS) was applied to the bilateral posterior motor cortices. Compared to sham stimulation, effective perturbation by rTMS slowed down the translation of sensorimotor-enriched L2 words—but not sensory-enriched L2 words. This finding suggests that sensorimotor-enriched training induced changes in L2 representations within the motor cortex, which in turn facilitated the translation of L2 words. The motor cortex may play a causal role in precipitating sensorimotor-based learning benefits, and may directly aid in remembering the native language translations of foreign language words following sensorimotor-enriched training. These findings support multisensory theories of learning while challenging reactivation-based theories.

**Keywords:** foreign language learning, motor cortex, multisensory, sensorimotor learning, TMS

## Introduction

A wealth of sensorimotor information is available for learning in natural environments. Sights, sounds, kinesthetic signals, and proprioceptive information may all be taken into account by learners as they acquire knowledge and skills. There is growing consensus that the brain appears optimized to function based on input arising across multiple sensorimotor modalities (Pasqualotto, Dumitru, & Myachykov, 2016; Röder & Wallace, 2010; Spence, 2018), and that training protocols incorporating sensorimotor functions can enhance learning (Meltzoff, Kuhl, Movellan, & Sejnowski, 2009; Shams, Wozny, Kim, & Seitz, 2011). We here refer to the presence of complementary sensory information during learning as *sensory enrichment* and the presence of complementary sensory and motor information during learning as *sensorimotor enrichment* (Mathias et al., 2019; Mayer, Yildiz, Macedonia, & von Kriegstein, 2015). Behavioral evidence that enriched training yields stronger learning outcomes relative to unisensory training has accumulated in several domains (Andrä, Mathias, Schwager, Macedonia, & von Kriegstein, 2020; MacLeod, Gopie, Hourihan, Neary, & Ozubko, 2010; Mathias, Palmer, Perrin, & Tillmann, 2015; for a review see Shams & Seitz, 2008; Sheffert & Olson, 2004; von Kriegstein et al., 2008).

Despite the potential for sensorimotor enrichment to serve as a powerful tool for learning (see for example Freeman et al., 2014), its underlying brain mechanisms remain largely unexplored. Research in the domain of foreign language (L2) learning has shown that cross-modal cortical responses occur following sensorimotor-enriched training: The translation of auditorily-presented L2 words into one's native language (L1) elicits pre-/motor cortex responses following gesture-based (sensorimotor-enriched) L2 learning (Macedonia & Mueller, 2016; Macedonia, Müller, & Friederici, 2011; Mayer et al., 2015). Moreover, behavioral benefits of sensorimotor-enriched L2 training have been shown to correspond to classification accuracy of a multivariate pattern analyzer (MVPA) trained to dissociate motor cortex responses to sensorimotor-enriched stimuli (Mayer et al., 2015). These findings suggest that motor cortex responses that occur during the translation of auditorily-presented

L2 words may be functionally linked to behavioral benefits of sensorimotor-enriched L2 training.

There are two opposing explanations regarding the relationship of responses within the motor cortices to behavioral benefits of sensorimotor-enriched training. On one hand, responses to unimodal stimuli within the motor cortices following sensorimotor-enriched learning may causally enhance stimulus recognition. This is the view taken by the predictive coding theory of multisensory learning (von Kriegstein, 2012), which proposes that sensory and motor cortices build up sensorimotor forward models during perception that simulate missing input (cf. Friston, 2012). On the other hand, responses to unimodal stimuli within sensory and motor brain regions following sensorimotor-enriched training may be viewed as epiphenomenal, a view taken by reactivation-based theories (Danker & Anderson, 2010; Fuster, 2009; Nyberg et al., 2001; Wheeler, Petersen, & Buckner, 2000). Reactivation theories suppose that motor brain responses engender a mere representation of a memorized stimulus, and therefore serve effectively no functional role in the recognition of the incoming stimulus. For example, if the sight of a bicycle triggered the recall of a motoric memory, reactivation theories assume that the recollected memory may arise from the reactivated motor cortex, but that the motor cortex responses would not aid in making the visual experience of the bike more precise.

Here, we tested the hypothesis that motor cortex causally contributes to benefits of sensorimotor-enriched training using transcranial magnetic stimulation (TMS). We perturbed a site within the motor cortex whose response patterns were previously found to correlate positively with the magnitude of sensorimotor-enriched learning benefits (Mayer et al., 2015). Noninvasive brain stimulation methods such as TMS permit to test whether stimulated brain areas casually contribute to ongoing behavioral outcomes, usually evidenced by increased response latencies (Day et al., 1989; Hartwigsen et al., 2017; Pascual-Leone, Wassermann, Grafman, & Hallett, 1996; Siebner, Hartwigsen, Kassuba, & Rothwell, 2009).

Our study had three aims. The first and primary aim was to test whether motor cortex causally contributes to the translation of recently-learned, sensorimotor-enriched L2 vocabulary. Based on the predictive coding theory of multisensory learning (von Kriegstein, 2012), we hypothesized that motor cortex stimulation would disrupt (i.e., slow down) the translation of sensorimotor-enriched L2 stimuli. Reactivation-based theories of learning (Fuster, 2009) do not consider differential effects of motor cortex stimulation on the translation of auditorily-presented L2 words. The second aim was to evaluate contributions of the motor cortex to the translation of sensorimotor-enriched L2 words over long post-training durations (e.g., several months). We expected greater effects of TMS on the translation of sensorimotor-enriched L2 words compared to picture-enriched vocabulary at later post-training time points. This hypothesis was based on evidence that effects of sensorimotor enrichment on L2 translation are stronger than effects of sensory enrichment half a year after learning the L2 vocabulary (Mayer et al., 2015). Our third aim was to compare effects of motor cortex stimulation on the translation of two different L2 word types: abstract and concrete nouns. We hypothesized that motor cortex stimulation would similarly influence the translation of both word types, based on previous results showing that sensorimotor enrichment can benefit the learning of both concrete and abstract words (Macedonia & Knösche, 2011; Macedonia, 2014; Mayer, Macedonia, & von Kriegstein, 2017).

## Materials and Methods

### Overview of the Experimental Design

The study consisted of a training phase followed by four TMS sessions, two immediately following the training phase and two as a long-term follow-up. During the training phase, native German speakers completed a four-day training program in which they learned L2 words and their L1 translations (**Figure 1a**). L2 vocabulary was learned in two conditions. In a sensorimotor-enriched learning condition, individuals viewed and performed gestures while L2 words were presented auditorily (gesture performance enrichment, **Figure 1b, left panel**). In a sensory-enriched learning condition, individuals viewed pictures while L2 words were presented auditorily (picture viewing enrichment, **Figure 1b, right panel**). We included these enrichment conditions in the current study for two reasons: First, of four enrichment conditions previously tested (gesture performance, gesture viewing, picture performance, picture viewing; Mayer et al., 2015), only these two conditions benefitted post-training L2 translation compared to auditory-only learning. Second, learning in these two conditions was associated with responses in distinct areas of the cerebral cortex (the motor cortex and the biological motion superior temporal sulcus for gesture performance enrichment and the lateral occipital complex for picture viewing enrichment; Mayer et al., 2015). For succinctness, we refer to the gesture performance enrichment simply as gesture enrichment, and picture viewing enrichment as picture enrichment.

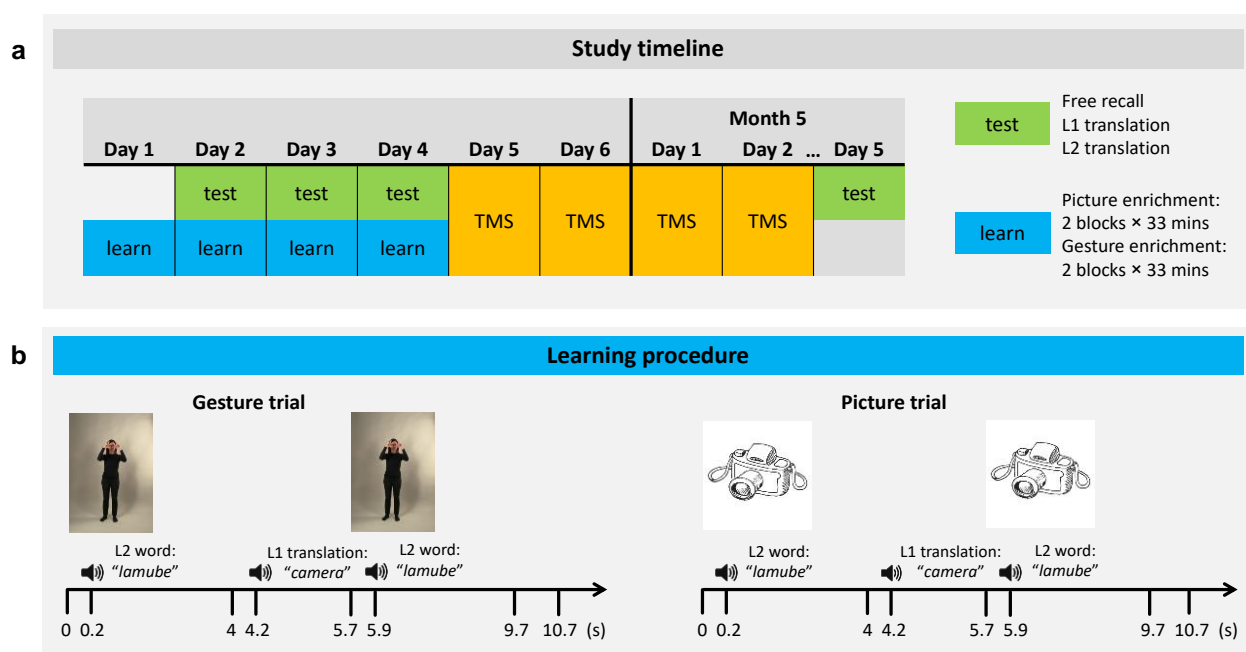

**Figure 1. Study timeline and foreign language learning procedure. (a)** Participants learned foreign language (L2) vocabulary over four consecutive days ('learn') in groups to emulate a classroom setting. Paper-and-pencil free recall and translation tests ('test') were administered on days 2 through 4 and approximately 5 months post-training. Transcranial magnetic stimulation ('TMS') sessions occurred on days 5 and 6 of the training phase, as well as on two consecutive days approximately 5 months post-training. Effective TMS was applied during one of the two sessions that occurred on days 5 and 6 and one of the two sessions that occurred in month 5, and ineffective sham stimulation was applied during the other sessions, with session orders counterbalanced across participants. **(b)** In both the gesture and picture enrichment conditions, participants heard an L2 word, followed by the translation in their native language (L1) and a repetition of the L2 word. Videos of iconic gestures and pictures accompanied L2 words in gesture and picture trials, respectively. Participants performed the gesture along with the video during its repetition. No motor component was added to the picture enrichment condition as this was previously found not to benefit L2 vocabulary learning (Mayer et al., 2015).

During the TMS sessions (**Figure 2a**), participants translated auditorily-presented L2 words into L1 while undergoing effective or sham TMS. The bilateral motor cortices were targeted with a combination of effective or sham offline TMS (TMS before the task; **Figure 2b**) and effective or sham online TMS (TMS during the task; **Figure 2c**). A combination of online and offline stimulation was used, as the right and left hemisphere motor sites were anatomically too close to each other for simultaneous dual-site online stimulation (see Hartwigsen et al., 2016 for a similar TMS design). Participants performed a multiple choice translation, in which they selected the correct L1 translation of an auditorily-presented L2 word from a list of options (**Figure 2c**); performance in this multiple choice task previously correlated with MVPA classification accuracy within the motor cortex site stimulated in the present study (Mayer et al., 2015). We also included an exploratory recall task in which participants pressed a button as soon as the L1 translation came to their mind when hearing each L2 word (results from this exploratory task are reported in the supplementary material). Participants did not perform gestures or view pictures during either of these tasks. Response time was used as the dependent variable because TMS typically influences response times rather than accuracy (Ashbridge, Walsh, & Cowey, 1997; Hartwigsen et al., 2017; Pascual-Leone et al., 1996; Sack et al., 2007). To evaluate long-term effects of motor cortex representations on enriched L2 vocabulary learning, TMS was administered both 5 days and 5 months following the start of training.

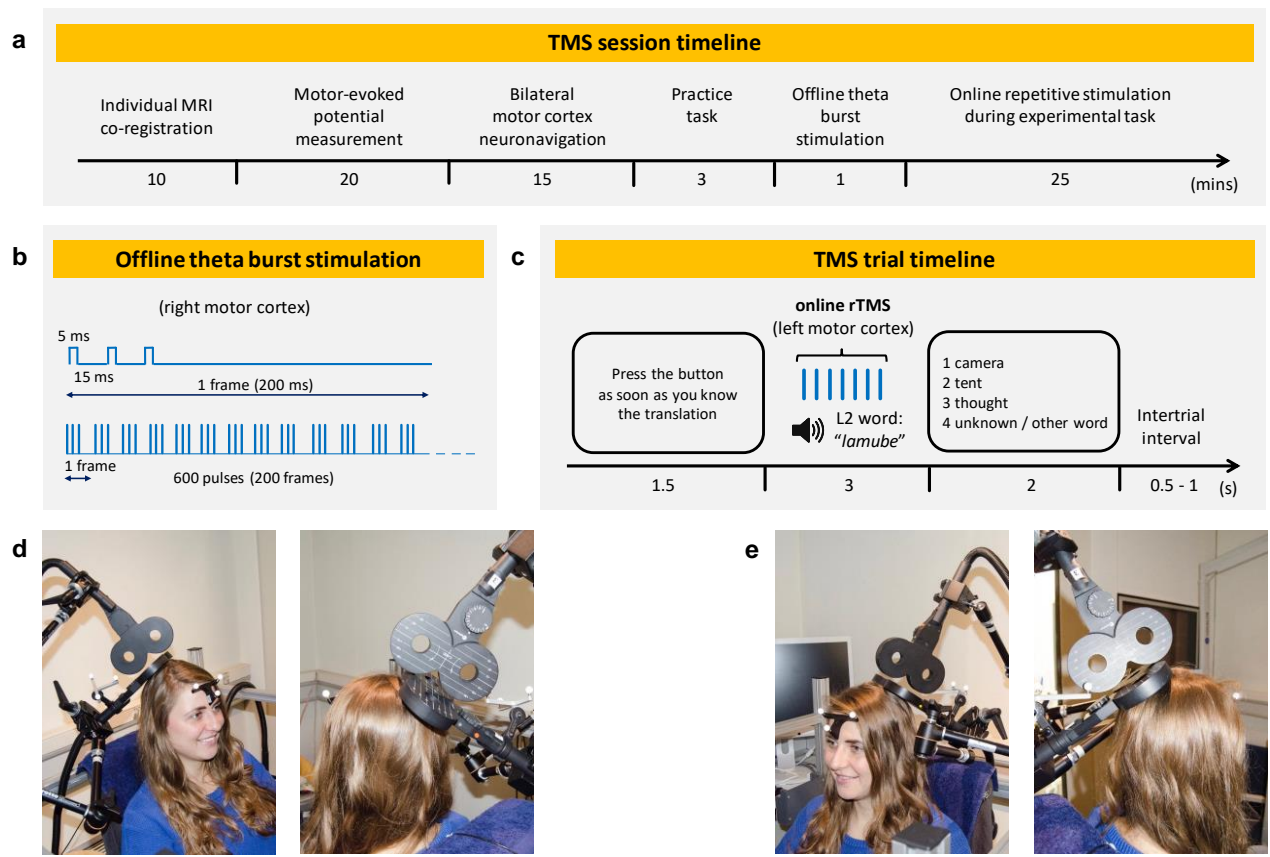

**Figure 2. Transcranial magnetic stimulation (TMS). (a)** Each TMS session began with the co-registration of the participant's scalp onto their previously-acquired T1-weighted structural MRI scan. Individual stimulation intensities were then acquired by measuring motor-evoked potentials. Stimulation coils were then neuronavigated to the bilateral motor cortex sites of interest for the experiment. Following a practice task, offline theta burst stimulation was delivered to the right motor cortex. The experimental TMS task was then performed as online repetitive TMS (rTMS) was delivered to the left motor cortex. Each TMS session lasted approximately 1.5 hours in total. **(b)** The offline theta burst stimulation consisted of 600 pulses, represented by blue lines, which were presented in 200-ms frames containing 3 pulses each. **(c)** On each trial, participants heard an L2 word that they had learned during the four-day training phase and pressed a button as soon as the L1 translation came to their mind (exploratory recall task). They then selected the L1 translation by button press from a list of options presented on a screen (multiple choice task). L1 words were presented in German. Online rTMS, which consisted of trains of seven TMS pulses at 10 Hz, was applied to the left motor cortex 50 ms following each L2 word onset. **(d)** Offline stimulation

was applied over the right motor cortex. **(e)** Online stimulation was applied over the left motor cortex.

## Participants

Twenty-two right-handed native German speakers completed the study (15 female; *M* age = 27.2 years, *SD* = 3.6 years). The sample size was based on two previous experiments (*n* = 22 per experiment) that estimated beneficial effects of gesture and picture enrichment on L2 learning outcomes (Mayer et al., 2015, Experiments 1 and 2). Participants were recruited from the in-house database of the Max Planck Institute for Human Cognitive and Brain Sciences in Leipzig, Germany and via an advertisement that was placed on the institute website. Of the 28 participants who registered for the study, one participant did not finish the four days of L2 training due to illness and withdrew from the study. Four other participants were excluded as they were unable to return for TMS sessions conducted 5 months after the training was completed, and one additional participant was excluded for studying the L2 vocabulary outside of the designated training hours.

Prior to the experiment, each participant was approved for TMS by a physician based on a safety screening questionnaire and physical examination. No participants reported a history of head injury, neurological, psychiatric, or psychological disorders, or contraindications for TMS. On average, the participants had studied one or more foreign languages for 14.3 years (*SD* = 6.9 years) and resided in a country in which a foreign language was spoken for 0.43 years (*SD* = 0.85 years). None of the participants were raised in bilingual homes. All participants reported normal or corrected-to-normal visual acuity; one participant who was weakly farsighted (right eye: +1.75 diopter (dpt), left eye: +2.25 dpt) reported being able to see the stimuli on a computer screen (approximately 1 m in front of the participant) during the TMS sessions without wearing glasses. No participants reported any hearing problems. Participants were compensated 212€ for completing the study. Written informed consent was obtained from each participant prior to beginning the experiment. The study was approved by the ethics committee of the University of Leipzig (*Aktenzeichen* 118/16-ek).

**Stimuli**

      **Vocabulary.** Participants learned 90 words selected from a corpus of pseudowords intended for experiments on foreign language learning and referred to as "Vimmi" (**Table 1**, Macedonia et al., 2010, 2011). The 90 words were randomly matched with 45 concrete and 45 abstract German nouns. Pseudowords were used for the purpose of avoiding potential associations with participants' previous L2 knowledge. The use of an artificial language also allowed us to control for several factors that potentially influence vocabulary learning, such as word length and frequency, which are difficult to control in a natural language. All L2 words consisted of three syllables and followed Italian phonotactic rules. Concrete and abstract German words were equivalent in terms of the number of syllables they contained (concrete $M$ = 2.40 syllables, $SD$ = 0.84; abstract: $M$ = 2.69 syllables, $SD$ = 0.90). Additionally, the frequency of concrete and abstract nouns in written German was roughly equivalent, according to the lexical database "*Wortschatzportal*" (http://wortschatz.uni-leipzig.de/de, concrete frequency scores: $M$ = 11.00, $SD$ = 1.18, range 9 to 13; abstract frequency scores: $M$ = 10.96, $SD$ = 0.98, range: 9 to 13).

| Concrete nouns | | | Abstract nouns | | |
|---|---|---|---|---|---|
| **German** | **English** | **Vimmi** | **German** | **English** | **Vimmi** |
| Ampel | traffic light | gelori | Absage | cancellation | munopa |
| Anhänger | trailer | afugi | Alternative | alternative | mofibu |
| Balkon | balcony | usito | Anforderung | requirement | utike |
| Ball | ball | miruwe | Ankunft | arrival | matilu |
| Bett | bed | suneri | Aufmerksamkeit | attention | fradonu |
| Bildschirm | monitor | zelosi | Aufwand | effort | muladi |
| Briefkasten | letter box | abota | Aussicht | view | gaboki |
| Decke | ceiling | siroba | Befehl | command | magosa |
| Denkmal | memorial | frinupo | Besitz | property | mesako |
| Eintrittskarte | entrance ticket | edafe | Bestimmung | destination | wefino |
| Faden | thread | kanede | Bitte | plea | pokute |
| Fahrrad | bicycle | sokitu | Disziplin | discipline | motila |
| Fenster | window | uribo | Empfehlung | recommenda- | giketa |
| Fernbedienung | remote control | wilbano | Gedanke | thought | atesi |
| Flasche | bottle | aroka | Geduld | patience | dotewa |
| Flugzeug | airplane | wobeki | Gleichgültigkeit | indifference | frugazi |
| Gemälde | painting | bifalu | Information | information | sapezo |
| Geschenk | present | zebalo | Korrektur | correction | fapoge |
| Gitarre | guitar | masoti | Langeweile | boredom | elebo |
| Handtasche | purse | diwume | Mentalität | mentality | gasima |
| Kabel | cable | zutike | Methode | method | efogi |
| Kamera | camera | lamube | Mut | bravery | wirgonu |
| Kasse | till | asemo | Partnerschaft | partnership | nabita |
| Katalog | catalog | gebamo | Rücksicht | consideration | ukowe |
| Kleidung | clothes | wiboda | Sensation | sensation | boruda |
| Koffer | suitcase | mewima | Stil | style | lifawo |
| Maschine | machine | nelosi | Talent | talent | puneri |
| Maske | mask | epota | Tatsache | fact | botufe |
| Papier | paper | serawo | Teilnahme | participation | pamagu |
| Reifen | tire | wasute | Tendenz | tendency | pefita |
| Ring | ring | guriwe | Theorie | theory | sigule |
| Rucksack | backpack | lofisu | Therapie | therapy | giwupo |
| Sammlung | collection | etuko | Tradition | tradition | uladi |
| Schlüssel | key | abiru | Triumph | triumph | gepesa |
| Schublade | drawer | lutepa | Übung | exercise | fremeda |
| Sonnenbrille | sunglasses | woltume | Unschuld | innocence | dafipo |
| Spiegel | mirror | dubeki | Veränderung | change | zalefa |
| Straßenbahn | tram | umuda | Verständnis | sympathy | gorefu |
| Tageszeitung | daily newspaper | gokasu | Vorgehen | procedure | denalu |
| Telefon | telephone | esiwu | Vorwand | excuse | pirumo |
| Teller | plate | buliwa | Warnung | warning | gubame |
| Teppich | carpet | batewo | Wohlstand | wealth | bekoni |
| Verband | bandage | magedu | Wohltat | benefaction | migedu |

**Table 1.** Vocabulary used in the experiment. 90 Vimmi and German words, and their English translations. Assignment of words to the gesture and picture enrichment conditions was counterbalanced across participants, ensuring that each Vimmi word was represented equally in both learning conditions.

**Audio files.** Audio recordings of the Vimmi and German words featured an Italian-German bilingual speaker who spoke with an Italian accent. Audio recording was conducted in a sound-damped chamber with a RØDE NT55 microphone (RØDE Microphones, Silverwater, Australia). Audio stimuli ranged in length from 654 to 850 ms ($M$ = 819.7 ms, $SD$ = 47.3 ms). Word lengths did not significantly differ between the subsets of words assigned to the two learning conditions, $t$ (88) = 1.30, $p$ = 0.20, or across the vocabulary types, $t$ (88) = 0.86, $p$ = 0.39. Examples of the audio stimuli are available at http://kriegstein.cbs.mpg.de/mayer_etal_stimuli/.

**Videos and pictures.** To conduct the language training, each L2 word was paired with a video or picture that illustrated the word's meaning (**Figure 1b**). Each video depicted an actress performing an iconic or symbolic gesture that embodied the meaning of a specific word. The actress was the same individual featured in the Vimmi and German audio recordings. Concrete words were enacted by iconic gestures that imitated their use or function (e.g., *window* was represented by the actress opening an imaginary window). Abstract words were embodied by symbolic gestures that illustrated their meaning (e.g., *plea* was represented by the actress taking a bow with her hands folded as in prayer). These symbolic gestures were previously agreed upon by three independent raters (Mayer et al., 2015). Each colored video was recorded by a Canon Legria HF S10 camcorder (Canon Inc., Tokyo, Japan) and lasted for 4 s. Videos always started and ended with the actress standing still with a neutral face in the center of the screen. All gestures consisted of head, uni- and bilateral arm, leg, or finger movements, or a combination of these movements. Movements were limited to a radius of one meter around the actress.

A professional graphic artist (https://www.klaus-pitter.com/) created a black-and-white line painting for each of the 90 L2 words (**Figure 1b**). Abstract words were often illustrated by complex scenes, while concrete words could usually be pictured as single objects. The complexity of visualization was not balanced between conditions in order to conform to differences that occur in natural learning settings.

Examples of the video and picture stimuli are available at http://kriegstein.cbs.mpg.de/mayer_etal_stimuli/.

## Design

The study had a 2 × 2 × 2 × 2 repeated-measures design. Within-participant independent factors were learning enrichment condition (gesture, picture), TMS condition (effective stimulation, sham stimulation), testing time point (5 days, 5 months), and L2 vocabulary type (concrete, abstract). The dependent variable was response time. We examined accuracy in the multiple choice TMS task in order to evaluate whether differences in response times between conditions could be attributed to speed-accuracy tradeoffs.

## Procedure

**Training phase.** Participants learned 45 L2 words in each of the two learning conditions. In the gesture enrichment condition, individuals viewed and performed gestures while L2 words were presented auditorily (**Figure 1b**). In the picture enrichment condition, individuals viewed pictures while L2 words were presented auditorily. To ensure that each of the 90 total L2 words was equally represented in both learning conditions, half of the participants learned one set of 45 words with gesture enrichment and the other 45 words with picture enrichment. The other half of participants received the reverse assignment of words to learning conditions in order to counterbalance the stimuli between participants and across learning conditions.

Each day of training comprised four 33-minute learning blocks. Each block consisted of 45 L2 words presented in a random order. Each word was repeated four times per block yielding a total of 180 trials per block. In two of the blocks, L2 words were enriched with their associated gestures, and in the other two blocks, L2 words were enriched with associated pictures. The order of learning blocks was counterbalanced across days and participant groups. Between blocks, the participants had breaks of 10 to 15 minutes. Snacks and drinks were provided for participants to consume during breaks.

Participants were instructed prior to the start of training that the goal was to learn as many L2 words as possible over the 4 days of training. Participants received no further instruction during the training except to be informed about which learning condition would occur next (i.e., gesture or picture enrichment). Since the L2 vocabulary learning took place in groups of up to 5 individuals, training sessions occurred in a seminar room with a projector and sound system. Audio recordings were played via speakers located on each side of the screen. The volume of the playback was adjusted so that all participants could comfortably hear the words. Participants were asked to stand during all learning blocks. The physical locations of individuals in each group within the training room were counterbalanced across training days.

Each gesture trial started with the presentation of a video showing a gesture for 4 s combined with the L2 word recording, which began 0.2 s after the video's onset (**Figure 1b**). The video was followed by a black screen paired with the audio presentation of the L1 translation for 1.7 s. Finally, the gesture video and corresponding L2 word were presented again for 4 s. The picture trials were similarly constructed. Each picture trial began with the presentation of the picture enrichment for 4 s combined with the recording of the L2 vocabulary, which began 0.2 s after the picture's onset. A black screen paired with the audio presentation of the L1 translation followed. Finally, the picture and corresponding L2 word were presented again. In the gesture-enriched blocks, participants were asked to perform the gestures along with the actress when the video was shown for the second time. A motor component was not included in the picture enrichment condition as a previous study found that the addition of a motor component to picture enrichment does not yield beneficial learning effects compared to non-enriched (auditory-only) learning (Mayer et al., 2015).

On days 2, 3, and 4 of the training phase, participants were asked to complete three paper-and-pencil vocabulary tests prior to beginning the training blocks. We included these tests in order to maintain the same L2 training procedure used by Mayer et al (2015).

Participants were given 15 minutes to complete each test. The first test was always a free recall test in which participants were asked to write down all L1 and L2 words, single or in combination, that they could remember. The second and third tests were L1-L2 and L2-L1 translation tests, in which participants were asked to translate lists of the 90 L1 or L2 words. The order of the two translation tests was counterbalanced across participants and training days. The order of words in the translation tests was randomized each day. The paper-and-pencil vocabulary test scores can be found in the supplementary material. The participant in each group with the highest score summed across the paper-and-pencil vocabulary tests administered on days 2, 3, and 4 was rewarded with an additional 21€ beyond the total study compensation. Participants were informed about this monetary incentive on day 1 prior to the start of the learning blocks.

Prior to beginning the training phase, all participants completed three psychological pretests that assessed their concentration ability (Concentration test, Brickenkamp, 2002) ($M$ score = 208.9, $SD$ = 34.3), speech repetition ability (Nonword Repetition test, Korkman, Kirk, & Kemp, 1998) ($M$ score = 101.5, $SD$ = 7.3) and verbal working memory (Digit Span test, Neubauer & Horn, 2006) ($M$ score = 18.2, $SD$ = 4.4). None of the participants were outliers (2 $SD$ above or below the group mean) with respect to their scores on any of the three tests, and all participants performed within the norms of the Concentration test for which norms were available.

**TMS sessions.** Following completion of the training phase, participants took part in the TMS sessions (**Figure 2a**) at two time points: 5 days and 5 months following the start of the training. At each time point, participants completed two TMS sessions. During one of these sessions, effective stimulation was delivered to the right and left posterior motor cortices, and during the other session, sham stimulation was delivered to the same sites.

Multiple choice and exploratory recall tasks completed during each of the four TMS sessions occurred in 4 blocks, and each block contained 45 trials, yielding a total of 180 trials per TMS session. Each of the 90 L2 words was therefore presented 2 times during each TMS session. Word orders were randomized within each block and for each participant. Participants responded during the translation test on a custom-made four-button response box using the index, middle, ring, and little fingers of the right hand. Each test trial began with the visually-presented instruction "Press the button as soon as you know the translation" ("*Drücken Sie den Knopf, sobald Sie die Übersetzung wissen*"), which was shown for 1.5 s on an EIZO 19" LCD monitor (EIZO Europe GmbH, Mönchengladbach, Germany; white letters, font: Arial, font size: 32 pt; black background). The screen was placed approximately 1 m in front of the participants. Next, one of the L2 words was presented over in-ear noise-cancelling headphones (Shure SE215. Shure Europe GmbH, Eppingen, Germany), combined with a black screen. The participants had 3 s, beginning at the L2 word's onset, to indicate by button-press if they knew the correct L1 translation of the auditorily-presented L2 word (exploratory recall task, **Figure 2c**). They were instructed to not press any button if they did not know the L1 translation. After 3 s, three possible L1 translations appeared on the screen, and participants selected the correct translation by pressing one of the four buttons on the response box (multiple choice task, **Figure 2c**). A fourth response option ("unknown / different word", "*unbekannt / anderes Wort*") was included in every trial. Participants were told to select this option if they did not know the correct L1 translation or if they had thought of a different L1 translation prior to the appearance of the three answer choices. The response options remained on the screen for 2 s, followed by a jittered inter-trial interval of 0.5 to 1 s ($M$ = 0.75 s). Participants were instructed to respond as quickly and as accurately possible.

The TMS sessions that were conducted 5 months after the start of the training phase followed the same procedure as the TMS sessions that occurred on days 5 and 6. The orders of effective and sham TMS sessions were counterbalanced across participants, both within each time point and between time points. The two TMS sessions were separated by a period of 18 weeks ($M$ = 18.2 weeks, $SD$ = 1.4 weeks).

Prior to beginning the translation task, participants completed a practice task with 20 common English words instead of Vimmi words. This practice task was administered during the first TMS session at each time point (i.e., 5 days and 5 months following the start of the training phase).

Finally, participants returned 1 to 6 days (*M* = 3.7 days, *SD* = 1.2 days) after their final TMS session, to complete again the three pencil-and-paper vocabulary tests (free recall, L1-L2 translation, and L2-L1 translation). Test orders were again counterbalanced across participants.

Participants had no knowledge of the month 5 follow-up TMS and behavioral sessions until they were contacted a few weeks prior to these month 5 testing dates. This was done to avoid potential rehearsal of the vocabulary during the 5-month interval between testing time points.

**Transcranial Magnetic Stimulation**

**Neuronavigation.** T1-weighted structural MRI scans were acquired for each participant using a 3-Tesla MAGNETOM Prisma-fit (Siemens Healthcare, Erlangen, Germany) with a  magnetization-prepared rapid gradient echo (MPRAGE) sequence in sagittal orientation (inversion time = 900 ms, repetition time = 2300 ms, echo time = 2.98 ms, flip angle = 9°, voxel-size = 1 ×1 ×1 mm). During the TMS sessions, the scalp of each participant was coregistered to their T1 scan using the neuronavigation software Localite (**Figure 2a**; Localite GmbH, Sankt Augustin, Germany).

Right and left posterior motor stimulation sites were based on fMRI results from a group of 22 participants (Mayer et al., 2015). These participants completed a similar sensorimotor-enriched L2 vocabulary training paradigm and a multivariate pattern classifier was trained to classify brain blood oxygen level-dependent (BOLD) responses to sensorimotor-enriched versus non-enriched auditorily-presented L2 words. TMS was applied over the coordinates that showed the highest positive correlation of the behavioral sensorimotor-enriched learning benefit with classifier accuracy (left hemisphere: *x, y, z* = −18, −43, 61 mm; right hemisphere: *x, y, z* = 21, −41, 73 mm, Montreal Neurological Institute (MNI) space) within the motor region of interest (ROI) used by Mayer and colleagues (2015). This ROI included the sum of the (bilateral) maps for Brodmann areas 4a, 4p, and 6 implemented in the Anatomy toolbox (Eickhoff et al., 2005) for SPM8 (Wellcome Trust Center for Neuroimaging, University College London, UK, http://www.fil.ion.ucl.ac.uk/spm/). MNI coordinates were transformed to individual space for each participant using SPM8.

**TMS parameters.** TMS pulses were generated by a MagPro X100 stimulator (MagVenture A/S, Farum, Denmark) and four focal figure-of-eight coils (C-B60; outer diameter = 7.5 cm). The pulses were triggered with Signal software version 1.59 (Cambridge Electronic Design Limited, Cambridge, UK). Presentation software (Neurobehavioral Systems, Inc., Berkeley, CA, www.neurobs.com) was used for stimulus delivery and response recording. Coils were held in position using fixation arms (Manfrotto 244, Cassola, Italy).

For offline TMS, we used continuous theta burst stimulation (cTBS), which consisted of 600 pulses divided into 200 frames (Huang et al., 2005; Valchev et al., 2017). Each frame had a duration of 200 ms and consisted of 3 separate square pulses with 5-ms durations and an inter-pulse-interval of 15 ms (**Figure 2b**). The cTBS frequency was 50 Hz and was delivered to the right hemisphere prior to the task. The online rTMS protocol consisted of seven single pulses delivered at 10 Hz beginning 50 ms after each stimulus onset (**Figure 2c**). Both TMS protocols are in line with the published safety guidelines (Rossi, Hallett, Rossini, & Pascual-Leone, 2009). Coils used in the sham TMS condition were positioned at a 90° angle over each stimulation coil (**Figure 2d** and **2e**), resulting in no effective brain stimulation. To avoid any potential carryover effects between TMS sessions, the sessions (i.e., the effective stimulation session and the sham stimulation session) were conducted on separate days (Rossi et al., 2009).

Effects of unilateral left hemisphere cTBS stimulation without rTMS were examined in a pre-study pilot experiment (see "Pre-study pilot experiment: Unilateral cTBS of left hemisphere motor cortex" in the supplementary material).

**Motor threshold measurement.** Motor thresholds were measured for each participant in order to determine individual stimulation intensities (**Figure 2a**). Single TMS pulses were applied to the left primary motor (M1) area controlling the first dorsal interosseous (FDI) muscle located on the right hand. A meta-analysis by Mayka and colleagues (2006) provided mean stereotactic coordinates of the left M1 ($x, y, z$ = −37, −21, 58 mm, MNI space), which were used as a starting point to locate the M1 FDI area.

Each participant's active motor threshold (AMT) was defined as the lowest stimulus intensity that caused at least five MEPs of 150 to 200 µV peak-to-peak amplitude out of ten consecutive TMS pulses in the tonically-active FDI muscle. A 50 µV peak-to-peak amplitude criterion was used to define the resting motor threshold (RMT) while the FDI muscle was relaxed. Following standard intensities used in previous studies, the cTBS stimulation intensity was set to 90% of AMT (Brückner, Kiefer, & Kammer, 2013; Chistyakov, Rubicsek, Kaplan, Zaaroor, & Klein, 2010), and rTMS intensity was set to 90% of the RMT (Halawa, Reichert, Sommer, & Paulus, 2019; Mathias et al., 2019; for a review see Thut & Pascual-Leone, 2010). The same cTBS and rTMS intensities were used for each individual across all four TMS sessions (cTBS *M* intensity = 36.4% of maximum stimulator output, *SD* = 5.9%; rTMS *M* intensity = 41.4% of maximum stimulator output, *SD* = 6.5%).

**Data Analysis**

Participants (*n* = 22) indicated that they recalled the L1 translation prior to the appearance of the four response options during roughly half of all trials across the two TMS sessions (*M* = 58.5% of trials, *SE* = 32.3%), leaving an insufficient number of trials for a meaningful analysis of this task component. An exploratory analysis of these data can be found in the supplementary results. For the multiple choice task, participants selected a translation from the options presented during *M* = 78.8% (*SE* = 21.3%) of trials across the two TMS sessions. In the following we therefore focus the analyses on response times for the multiple choice task.

To test our first hypothesis (see the last paragraph of Introduction), we ran a two-way repeated measures ANOVA with the factors learning condition (gesture, picture) and stimulation type (TMS, sham) on the multiple choice task response times. To evaluate whether the observed patterns of response times were due to speed-accuracy tradeoffs, we correlated response times in the multiple choice task with accuracy (percent correct) for each learning and stimulation condition.

To test our second hypothesis on differential contributions of the motor cortex at early and later time points, we ran a three-way repeated measures ANOVA on multiple choice task response times with factors learning condition (gesture, picture), stimulation type (effective, sham), and time point (day 5, month 5). We followed up the three-way ANOVA with two-way repeated measures ANOVAs on multiple choice task response times with factors learning condition (gesture, picture) and stimulation type (effective, sham) for each time point.

To test our third hypothesis regarding influences of vocabulary type on multiple choice task response times, we ran three-way repeated measures ANOVAs with factors learning condition (gesture, picture), stimulation type (effective, sham), and vocabulary type (concrete, abstract) on response times at each time point.

Pairwise comparisons were conducted using Tukey HSD post-hoc tests. $\eta_p^2$ and Hedge's *g* were used as measures of effect size.

Multiple choice task accuracy, as well as paper-and-pencil vocabulary test scores (free recall and translation tests), are reported in the supplementary material (**Tables S1 and S2**).

## Results

**Motor cortex stimulation slows the translation of gesture-enriched foreign vocabulary**

Our first and primary hypothesis was that motor cortex integrity would contribute to L2 translation following gesture-enriched L2 learning, but not picture-enriched L2 learning. We therefore first tested whether motor cortex stimulation modulated L2 translation, irrespective of testing time point and vocabulary type, but depending on learning condition. The results confirmed our hypothesis. A two-way analysis of variance (ANOVA) on response times in the multiple-choice task revealed a stimulation type × learning condition interaction ($F_{1,21} = 4.15$, $p = .04$, two-tailed, $\eta_p^2 = .16$). Tukey's HSD post-hoc tests revealed that response times for words that had been learned with gesture enrichment were significantly delayed when TMS was applied to the motor cortex compared to sham stimulation ($q$ = 3.9, $p$ = .04, Hedge's $g$ = .17). TMS and sham conditions did not significantly differ for words learned with picture enrichment ($p$ = .94), indicating that perturbation of a brain area related to motor function slowed the translation of L2 words that had been learned with gestures, but not of L2 words learned with pictures (**Figure 3**).

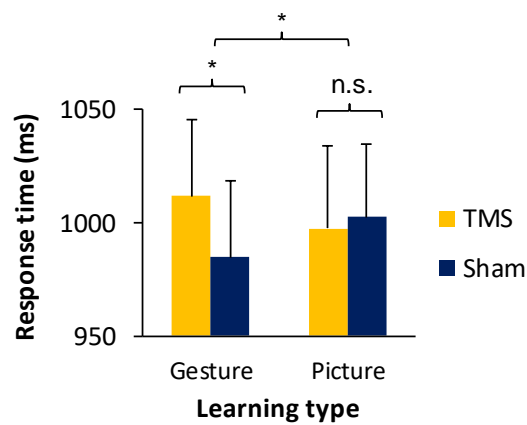

**Figure 3. Effects of motor cortex stimulation on L2 translation in the multiple choice task.** Bilateral motor cortex stimulation slowed the translation of L2 vocabulary learned using gestures compared to sham stimulation. There was no such effect on L2 vocabulary learned using pictures. The mean of each condition across time points (5 days and 5 months following the start of learning) is shown ($n$ = 22 participants). Error bars represent one standard error of the mean. *$p$ < .05.

In a control analysis, we tested whether differences in response times under effective stimulation compared to sham stimulation conditions could be due to tradeoffs between translation speed and accuracy. Response times for correct responses in the multiple choice translation task were compared with accuracy (percent correct) for each learning and stimulation condition. Response times correlated negatively with translation accuracy: Slower responses were associated with lower accuracy for all conditions (gesture-TMS, $r$ (20) = −.62, $p$ = .002, Bonferroni corrected; gesture-sham, $r$ (20) = −.54, $p$ = .009; picture-TMS, $r$

(20) = −.60, *p* = .003, picture-sham, *r* (20) = −.55, *p* = .008; **Figure 4**). The correlations for TMS and sham conditions did not significantly differ (all *p*'s > .36). These results suggest that participants did not trade speed for accuracy in the multiple choice task.

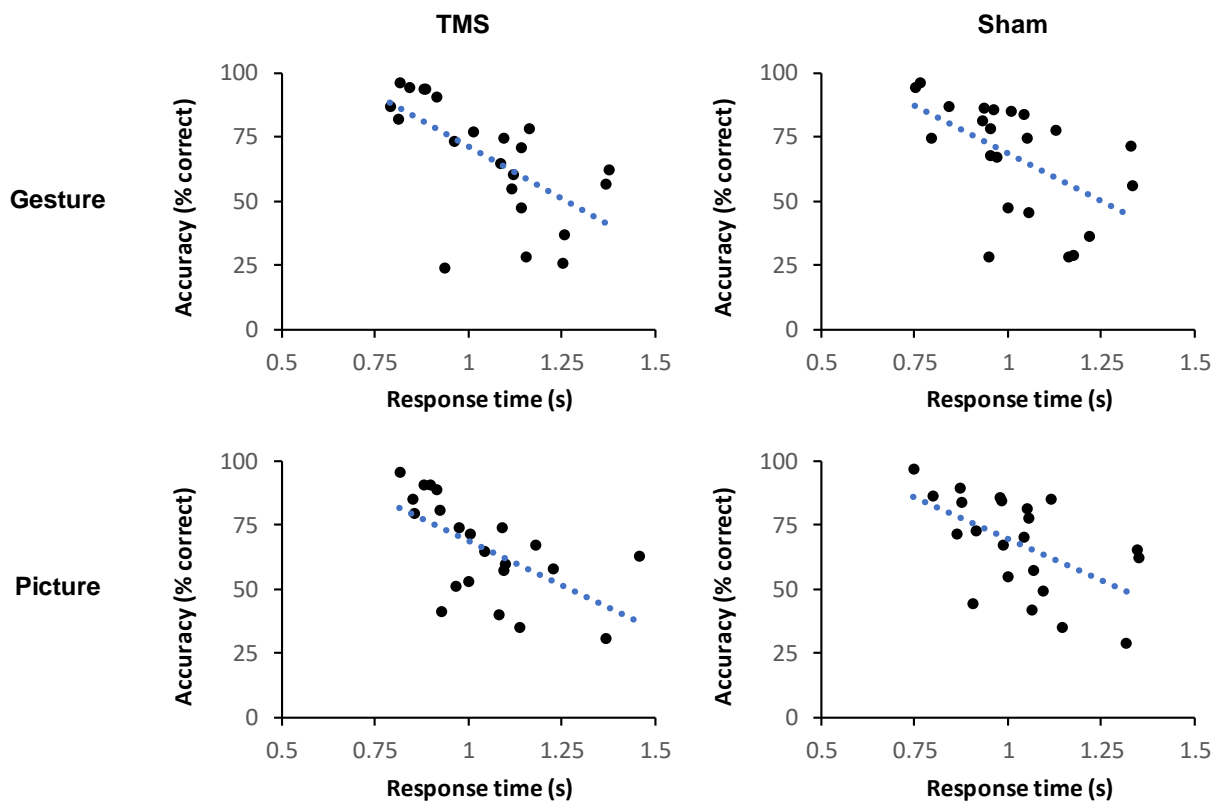

**Figure 4. Speed-accuracy relationships in L2 translation in the multiple choice task.** Slower response times correlated with lower translation accuracy, indicating that there was no speed-accuracy tradeoff in the multiple choice TMS task. All *p*'s < .01, Bonferroni corrected; *df* = 20 for all correlations.

**Effects of motor cortex stimulation occurred 5 days following the start of L2 training**

We next tested our second hypothesis that effects of motor cortex stimulation on the translation of gesture-enriched L2 words would be greater at the later post-training time point (month 5) compared to the earlier post-training time point (day 5). Contrary to our hypothesis, a three-way ANOVA with the factors stimulation type, learning condition, and time point on translation response times in the multiple-choice task yielded no significant three-way interaction. There was a significant main effect of time point. Participants responded faster at day 5 than month 5 ($F_{1, 21}$ = 62.21, $p$ < .001, two-tailed, $\eta_p^2$ = .75). There were no other main effects or interactions.

To explore whether the stimulation type and learning condition variables interacted significantly within each time point, we performed separate two-way ANOVAs with the factors stimulation type and learning condition on response times at each time point. The ANOVA on day 5 response times yielded a significant stimulation type × learning condition interaction ($F_{1, 21}$ = 4.59, $p$ = .04, two-tailed, $\eta_p^2$ = .18). Tukey's HSD post-hoc tests revealed that responses were significantly slower for the gesture condition during the application of TMS compared to sham stimulation ($q$ = 3.9, $p$ = .04, Hedge's $g$ = .16). The ANOVA on month 5 response times yielded no significant main effects or interaction.

In sum, significant effects of motor cortex stimulation on L2 translation occurred 5 days following the start of the L2 training period, and, although effects of stimulation were not significant at month 5, there was no significant interaction of TMS effects between the two time points (**Figure 5**).

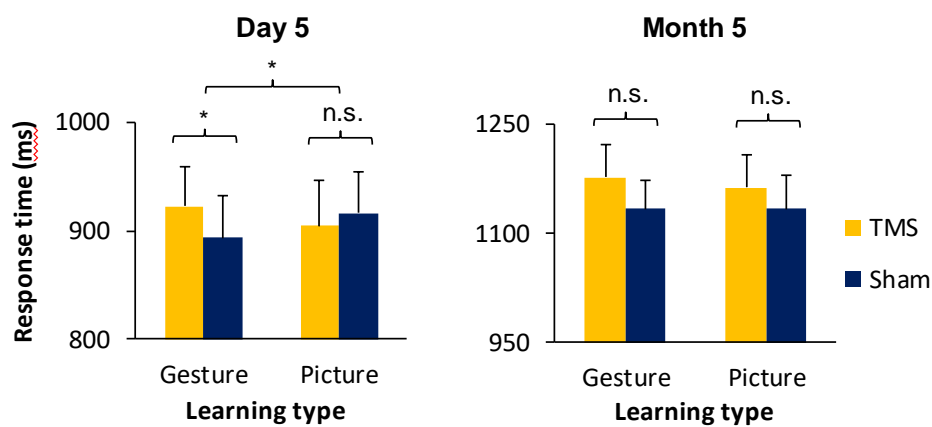

**Figure 5. Effects of motor cortex stimulation on L2 translation by time point in the multiple choice task.** Effects of motor stimulation occurred 5 days following learning ($n$ = 22 participants). There was no significant three-way interaction between time point, stimu­lation type, and learning condition factors. Error bars represent one standard error of the mean. *$p$ < .05.

**Motor cortex integrity influences L2 translation independent of word type**

      Finally, we tested the hypothesis that the motor cortex integrity would influence L2 translation independent of word type (i.e., whether a word was concrete or abstract). We conducted a four-way ANOVA on translation response times in the multiple choice task with the factors stimulation type, learning condition, time point, and vocabulary type. The four-way ANOVA yielded a two-way stimulation type × learning condition interaction ($F_{1,\ 21}$ = 5.20, $p$ = .03, two-tailed, $\eta_p^2$ = .20). Response times for words that had been learned with gesture enrichment were significantly delayed when TMS was applied to the motor cortex compared to sham stimulation ($q$ = 3.9; $p$ = .007, Hedge's $g$ = .19). As expected, the stimulation type × learning condition interaction was not influenced by the word type variable: the three-way stimulation type × learning condition × word type interaction was not significant. There was also no significant four-way interaction (**Figure 6**). Thus, effects of neurostimulation on gesture-enriched words were not modulated by word type.

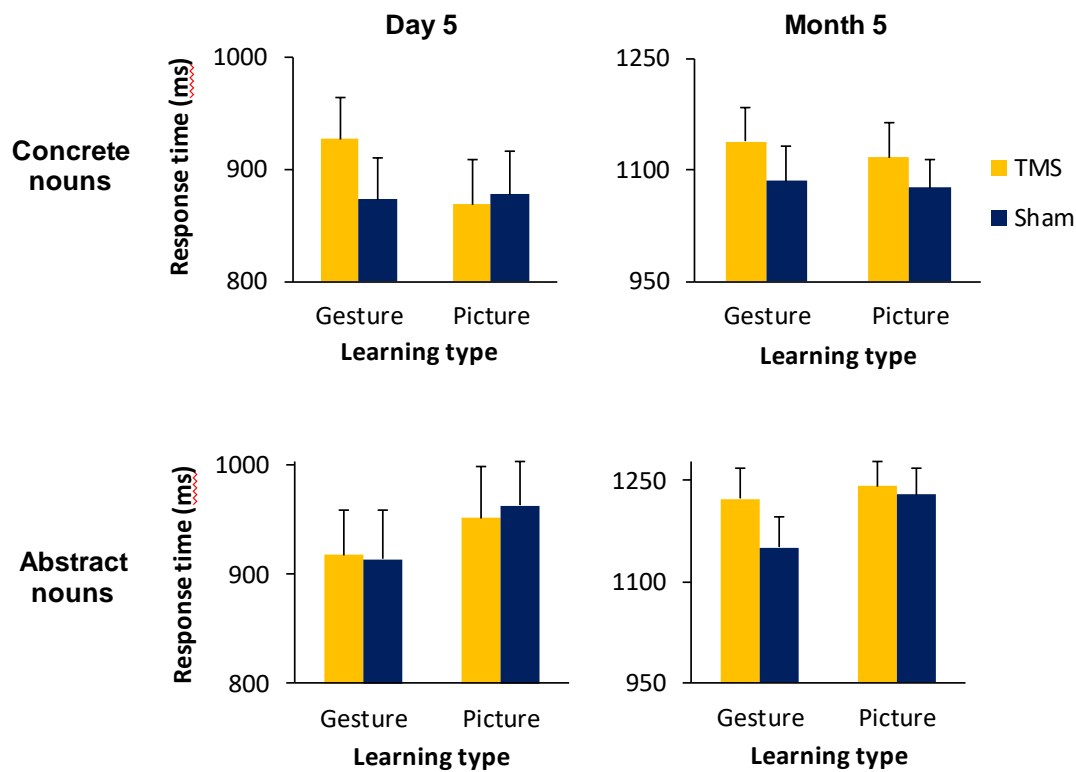

**Figure 6. Effects of motor cortex stimulation on L2 translation by time point and vocabulary type in the multiple choice task.** L2 vocabulary translation response times at the day 5 TMS session (left) and month 5 TMS session (right) by stimulation type, learning type, and word type (*n* = 22 participants). Error bars represent one standard error of the mean.

There was a main effect of vocabulary type ($F_{1, 21}$ = 43.52, *p* < .001, two-tailed, $\eta_p^2$ = .67): Concrete words were translated more rapidly than abstract words at both time points. This effect was expected based on previous studies (Macedonia & Klimesch, 2014; Macedonia & Knösche, 2011). This effect was qualified, however, by a significant time point × vocabulary type interaction ($F_{1, 21}$ = 7.02, *p* = .02, two-tailed, $\eta_p^2$ = .25). Abstract words took significantly longer to translate than concrete words at month 5 (*p* = .003, Hedge's *g* = .52), while abstract and concrete word translation times did not significantly differ at day 5.  There was also an expected main effect of time point ($F_{1, 21}$ = 64.08, *p* < .001, two-tailed, $\eta_p^2$ = .75):

L2 words were translated more rapidly at day 5 than month 5. Finally, there was an unexpected learning condition × vocabulary type interaction ($F_{1, 21}$ = 17.15, $p$ < .001, two-tailed, $\eta_p^2$ = .45). Picture-enriched abstract words took significantly longer to translate than picture-enriched concrete words ($p$ = .003, Hedge's $g$ = .48), while the time it took to translate concrete and abstract gesture-enriched words did not significantly differ.

## Discussion

We used inhibitory TMS to test the causal relevance of motor cortex function following sensorimotor-enriched L2 vocabulary learning. There were three main findings. First, motor cortex stimulation disrupted the translation of recently-learned, gesture-enriched vocabulary relative to sham stimulation. This result supports our principal hypothesis that motor cortex integrity causally contributes to the translation of sensorimotor-enriched L2 vocabulary. Motor cortex stimulation had no effect on the translation of picture-enriched vocabulary. Second, motor cortex integrity benefitted L2 translation in the short-term but not, contrary to our hypothesis, in the long-term. Third, in congruence with our hypothesis, effects of motor cortex stimulation on the translation of sensorimotor-enriched L2 vocabulary were not influenced by whether the L2 word was concrete (e.g., *ball*) or abstract (e.g., *thought*). Taken together, these findings suggest that sensorimotor-enriched training constructs associations between L2 words and their L1 translations by way of representations arising from the motor cortex.

Motor cortex integrity influenced task performance in another sensory modality, depending on associations forged during learning. This finding differentiates between multisensory predictive coding (Friston, 2012; von Kriegstein, 2012) and reactivation-based (Danker & Anderson, 2010; Fuster, 2009) theories of learning. The causal relation observed between the motor cortex and auditory L2 translation is predicted by multisensory predictive coding theories, but not by reactivation theories. What functional mechanisms may explain the involvement of the motor cortex in beneficial effects of sensorimotor-enriched learning?

Predictive coding theories suggest that, in order to recall the meaning of a newly-acquired L2 word, one may internally re-enact or simulate the perceptual and motor processes that were involved in learning that word via a generative model (Friston & Kiebel 2009; Rao & Ballard, 1999; von Kriegstein et al., 2008). Numerous behavioral studies support the notion that perceptual and motor processes involved in learning are re-enacted during recognition (Dudschig, Souman, Lachmair, Vega, & Kaup, 2013; Estes, Verges, & Barsalou, 2008; Meade, Midgley, Sehyr, Holcomb, & Emmorey, 2017; Matheson, Familiar, & Thomson-Schill, 2018; Witt, Kemmerer, Linkenauger, & Culham, 2010; Zwaan & Yaxley, 2003). During reading, for example, concrete nouns such as *plane* and *foot* elicit upward and downward saccades, respectively, suggesting that visuomotor simulations are involved in accessing semantic knowledge (Estes et al., 2008; replicated in Dudschig et al., 2013). In the present study, the motor cortex may have supported the internal simulation of movement-related information during auditory L2 translation, thereby benefitting response times. This, however, remains speculative, as the present study was not designed to conclusively demonstrate that a predictive coding mechanism underlies the causal relevance of motor cortex representations in L2 translation.

Several alternative theories cannot fully account for the causal engagement of motor cortex in L2 translation following sensorimotor-enriched training. Levels-of-processing theory (Craik & Lockhart, 1972) attributes enhanced memory performance in word learning tasks to the processing of word meaning across a greater number of levels (e.g., a semantic level and an orthographic level), associated with increased activity in prefrontal and temporal areas (Nyberg, 2002). Other lines of work have pointed towards the efficacy of emotion- or attention-based interventions for performance on cognitive tasks, such as the positive effects of background music on IQ tests (Schellenberg, Nakata, Hunter, & Tamoto, 2007). However, if behavioral performance following enrichment-based learning was determined solely by levels-of-processing or emotion-based mechanisms, then stimulation of a motor area would not have disrupted L2 translation. Another theoretical explanation is that sensorimotor and sensory enrichment affects learning outcomes by influencing brain responses in the sensory modality experienced at test, e.g., the auditory modality for auditory tests (for a review see Shams &

Seitz, 2008). This account, also, would have predicted no influence of motor cortex integrity on L2 translation following gesture-enriched learning. In sum, these alternative mechanisms are unable to explain how motor cortex integrity contributed to post-enrichment performance. Educational strategies may advance by establishing links between to-be-learned information and sensorimotor and sensory enrichment (cf. Markant, Ruggeri, Gureckis, & Xu, 2016).

It is likely that neural representations of enrichment-related information during L2 translation are not limited to only motor-related information, but rather extend to additional relevant input received during enriched L2 learning. Support for this inference comes from a recent study showing that inhibitory stimulation of the biological motion superior temporal sulcus (bmSTS), a region specialized in the processing of visual biological motion (Grossman & Blake, 2001; Jastorff, Kourtzi, & Giese, 2006), also disrupted the auditory translation of gesture-enriched (but not picture-enriched) L2 words (Mathias et al., 2019). This view is consistent with the idea that the presence of an additional dimension (e.g., motor, visual) along which stimuli can be evaluated during recognition underlies the beneficial effects of sensorimotor-enriched learning (Liu & Wang, 2018). Neural representations of sensory and motor input experienced during learning may also include a somatosensory component. Incidentally, the motor area stimulated in the current study, which was identified by Mayer and colleagues (2015) as the region correlating most strongly with behavioral performance following sensorimotor-enriched L2 learning, was spatially proximal to the somatosensory cortex. Similar patterns of activity occurring at the boundary of motor cortex and the somatosensory cortex have been observed following children's learning of symbols through handwriting (Vinci-Booher, James, & James, 2016). It is well-established that one's own body movements are strongly associated with memories of their sensory outcomes (Hommel, Brown, & Nattkemper, 2016), and that memory for actions, therefore, may rely not only on motor representations, but also on representations of the sensory consequences of actions (Badets & Osiurak, 2015). Given that action-based enrichment techniques are consistently accompanied by sensory feedback, and that perceptual and motor learning appear reciprocally linked and generally occur together (for a review see Ostry & Gribble, 2016), it is likely that neural representations of enrichment-related information include both motor and associated somatosensory components. The magnetic field that was used to stimulate the motor cortex may have also impacted some part of nearby somatosensory cortex, as the effective area of figure-of-eight coil stimulation is approximately 1 to 2 cm (Sandrini, Umiltà, & Rusconi, 2011; Walsh & Rushworth, 1999); increasing the focality of stimulation to a smaller cortical target is not currently possible, except through significantly more invasive stimulation methods such as the placement of electrodes on the cortical surface.

The perception of familiar vocabulary in the native language activates an experience-dependent network of sensory and motor areas (for reviews see Kiefer & Pulvermüller, 2012; Lambon Ralph, Jeffries, Patterson, & Rogers, 2017) and enhances motor cortex excitability (Watkins, Strafella, & Paus, 2003). While some have argued that motor activity during L1 perception is merely a downstream by-product of L1 processing in "core" language areas (Lotto, Hickok, & Holt, 2009), other studies have provided evidence that motor structures contribute to speech perception (Rogers, Möttönen, Boyles, & Watkins, 2014; Stokes, Venezia, & Hickok, 2019; for a review see Skipper, Devlin, & Lametti, 2017) or have argued that perceiving speech is essentially perceiving gestures (motor theory of speech perception; Liberman & Mattingly, 1985; for a review see Galantucci, Fowler, & Turvey, 2006). Motor structures may functionally support L1 comprehension by representing conceptual aspects of language processing (for reviews see Barsalou, 2005; Kemmerer, 2015; Kemmerer, Rudrauf, Manzel, & Tranel, 2012; Pulvermüller, 2012); concepts may be implemented via a predictive coding mechanism (for a review see Matheson & Barsalou, 2018). Hearing L1 words that refer to body movements such as *pick* and *kick* (Hauk, Johnsrude, & Pulvermüller, 2004), as well as action words related to the use of tools (Marino, Borghi, Buccino, & Riggio, 2017) can trigger responses in premotor and motor cortices. Neurostimulation evidence has pointed to the functional relevance of motor areas in behavioral responses to L1 words that refer specifically to body movements (Vukovic, Feurra, Shpektor, Myachykov, & Shtyrov, 2017). Our findings add a fundamentally novel line of research to these previous results by demonstrating the causal relevance of motor responses in the representation of recently-acquired sensorimotor-enriched stimuli, and, in particular, non-action words such as concrete and abstract nouns. The use of an artificially-generated vocabulary in the current study permitted the control of variables related to participants' learning experience, as well as stimulus-related variables such as word lengths, phonological contents, functions, speaker, and so forth. Thus, the current findings suggest that unfamiliar words can become associated with information represented within the motor cortex through short-term sensorimotor-enriched learning, and that this information is able to causally influence the auditory translation of those words.

Contributions of the motor cortex to the translation of L2 vocabulary arose following a relatively brief period (four days) of gesture-enriched L2 training. The short training period required suggests that sensory and motor elements of sensorimotor-enriched learning experience can be integrated rapidly with vocabulary representations, and that lifelong experience with words and their semantic associations is not a prerequisite for establishing representations of auditory words within motor cortex. The relatively rapid acquisition of multisensory representations has been demonstrated in some previous studies. Thelen and Murray (2012) report effects of single-trial audiovisual learning on evoked potentials elicited during subsequent visual-only recognition. von Kriegstein and Giraud (2006) demonstrated that two minutes of exposure to a voice paired with a face enhanced functional coupling between voice and face brain areas during subsequent perception of the trained voices. In another study, multisensory audiovisual training on a low-level visual task resulted in significantly faster learning than unisensory training over a ten-day period (Seitz, Kim, & Shams, 2006). Effects of motor cortex stimulation on the translation of sensorimotor-enriched L2 words in the current study were limited to the day immediately following the four-day training program. This result was not expected, as effects of sensorimotor enrichment on L2 translation have been shown to exceed those of sensory enrichment up to several months following enriched L2 vocabulary training (Mayer et al., 2015). One explanation for the differential effects of stimulation at the two time points tested here could be that participants in the current study received a lesser amount of training compared to those in the study conducted by Mayer and colleagues (2015), potentially diminishing the magnitude of enrichment-based effects at the later time point.

We conclude that behavioral performance in vocabulary translation following sensorimotor-enriched training is supported at least in part by representations in the motor cortex. The translation of recently-acquired foreign language words may therefore rely not only on auditory information stored in memory or modality-independent L2 representations, but also on the sensorimotor context in which the L2 words have been experienced. The causal relation observed between motor cortex stimulation and behavioral performance contributes to the broader neuroscientific debate on the role of motor brain structure in human cognitive functions; specialized sensorimotor brain mechanisms contribute beneficially to sensorimotor-enriched learning.

## Acknowledgements

This work was funded by the German Research Foundation grant KR 3735/3-1, a Max Planck Research Group grant to K.v.K, and an Erasmus Mundus postdoctoral fellowship in Auditory Cognitive Neuroscience. B.M. is also supported by the ERC-Consolidator Grant SENSOCOM 647051 to K.v.K.

**Supplementary Material**

**Pre-study pilot experiment: Unilateral cTBS of left hemisphere motor cortex**

      **Introduction.** In a pilot experiment conducted prior to the current study, we explored whether unilateral left hemisphere stimulation using continuous theta burst stimulation (cTBS) is sufficient to interfere with L2 vocabulary translation for gesture enriched words. We tested effects of unilateral left-lateralized motor stimulation because, in the previous functional magnetic resonance imaging (fMRI) study on enriched L2 translation (Mayer et al., 2015), multivariate pattern analyzer (MVPA) accuracy significantly correlated with the behavioral benefits of gesture-enriched learning within the left motor cortex, while the same test was not significant in the right hemisphere motor cortex. Several recent studies, however, have demonstrated null effects of unilateral compared to bilateral TMS (Jelić, Filipović, Milanović, Stevanović, & Konstantinović, 2017; Park et al., 2017; Ritzinger, Huberle, & Karnath, 2012; Yang et al., 2018).

      **Participants.** Eleven participants were recruited from the in-house database of the Max Planck Institute for Human Cognitive and Brain Sciences in Leipzig, Germany and via an advertisement that was placed on the institute website. None of these participants also took part in the study reported in the main manuscript. The study inclusion criteria were the same as those reported for the main experiment (right-handed native German speakers, normal or corrected-to-normal visual acuity, and so forth). One participant withdrew from the pilot experiment for a medical reason during the four-day training phase that preceded the TMS sessions, leaving 10 remaining participants (6 female; *M* age = 25.6 years, *SD* = 4.2 years) for analysis.

      **Design, procedure, and stimuli.** The pilot experiment followed the same repeated-measures design and procedure as the study reported in the main manuscript (**Figure 1a**), with the exception that TMS was delivered only on days 5 and 6 (and not 5 months following the training phase). The within-participant independent factors were therefore learning enrichment condition (gesture, picture), TMS condition (effective stimulation, sham stimulation), and L2 vocabulary type (concrete, abstract). The dependent variable was

response time in the TMS tasks. The pilot experiment used the same 90 L2 stimuli as the study reported in the main manuscript (**Figure 1b**).

**Results and discussion.** The hypothesized learning condition × stimulation type interaction did not reach significance in the two-way repeated measures ANOVA with factors learning condition (gesture, picture) and stimulation type (effective, sham) on response times in the multiple choice task ($F_{1, 9}$ = .41, $p$ = .54, two-tailed, $\eta_p^2 = $ .04). We therefore decided to stimulate the motor cortex bilaterally in the current study. A combination of offline cTBS applied to the left motor cortex and online repetitive TMS (rTMS) applied to the right motor cortex was used because the neuronavigated target coordinates were spatially too close to the midline to permit the concurrent placement of two coils on the scalp (see Hartwigsen et al., 2016 for a similar TMS design). Effects of offline cTBS on neural motor function last for a duration of nearly 60 minutes (Huang, Edwards, Rounis, Bhatia, & Rothwell, 2005), while effects of online high-frequency rTMS inhibit motor function for a period lasting for about half the duration of the stimulation train (Rotenberg, Horvath, & Pascual-Leone, 2016). Thus, we hypothesized that the combination of offline cTBS and online rTMS would yield bilateral inhibitory effects on motor cortex function in the current study.

**Analysis of response times in the exploratory recall task**

In the exploratory recall task, response time was defined as the time elapsed between the start of the auditory L2 word presentation and the participant's indication by button press (prior to the appearance of the four response options) that they knew the L1 translation of the presented L2 word. Participants indicated that they recalled the L1 translation prior to the appearance of the four response options during roughly half of all trials across the two TMS sessions (*M* = 58.5% of trials, *SE* = 32.3%), leaving an insufficient number of trials for analysis of this task component. We nevertheless performed an exploratory analysis of recall response times. We analyzed trials in which participants first indicated by button press that they recalled the L1 translation and subsequently selected the correct translation from the list of response options presented on the screen.

A four-way ANOVA on recall response times for correct response trials with factors learning condition, stimulation type, time point, and vocabulary type yielded a significant main effect of time point, ($F_{1, 21} = 42.34$, $p < .001$, two-tailed, $\eta_p^2 = .67$). Recall response times were significantly faster at day 5 than month 5. There was, however, no significant main effect of vocabulary type ($p = .30$). Recall response times for concrete words ($M = 1402$ ms, $SE = 18$ ms) did not differ from response times for abstract words ($M = 1426$ ms, $SE = 22$ ms). The ANOVA yielded a significant learning condition × vocabulary type interaction ($F_{1, 21} = 4.72$, $p = .04$, two-tailed, $\eta_p^2 = .18$). The predicted two-way interaction between learning condition and stimulation type variables was not significant ($p = .60$). Response times did not significantly differ between any conditions (TMS-Gesture: $M = 1410$ ms, $SE = 30$ ms; Sham-Gesture: $M = 1409$ ms, $SE = 28$ ms; TMS-Picture: $M = 1431$ ms, $SE = 32$ ms; Sham-Picture: $M = 1405$ ms, $SE = 24$ ms). There were no other significant main effects or interactions.

Given that not even the usually robust difference between concrete and abstract vocabulary types (Atkinson & Juola, 1973; Macedonia & Knösche, 2011; Paivio, 1965; Walker & Hulme, 1999) emerged in this analysis of recall response times, we assume that the low response rate yielded an insufficient number of trials for analysis of this task component. An alternative interpretation is that there was no effect of motor cortex stimulation on this particular task.

**Table S1.** Response accuracy in the TMS multiple choice task. *M* = mean accuracy (% correct), *SE* = standard error.

| | | | Stimulation type | |
|---|---|---|---|---|
| **Learning type** | **Time point** | **Vocabulary type** | TMS *M* (*SE*) | Sham *M* (*SE*) |
| Gesture | Day 5 | Abstract | 80 (4.7) | 80 (4.8) |
| | | Concrete | 84 (4.5) | 83 (4.5) |
| | Month 5 | Abstract | 48 (6.1) | 50 (5.7) |
| | | Concrete | 57 (5.8) | 58 (5.0) |
| Picture | Day 5 | Abstract | 77 (4.5) | 76 (4.7) |
| | | Concrete | 86 (2.7) | 86 (3.3) |
| | Month 5 | Abstract | 44 (5.6) | 46 (5.1) |
| | | Concrete | 58 (5.2) | 64 (4.8) |

**Paper-and-pencil tests**

Mean paper-and-pencil vocabulary test scores are shown in **Table S2**. Paper-and-pencil tests were independently scored for accuracy by two raters. Free recall were scored in terms of the number of translations (German-Vimmi or Vimmi-German word pairs), German words that were missing corresponding Vimmi words, and Vimmi words that were missing corresponding German words. Three points were given for each correct translation (German-Vimmi or Vimmi-German word pair). One point was given for each correctly-recalled German word that was missing a corresponding Vimmi translation and vice versa. Points were summed to determine a total score for each participant, learning condition, and vocabulary type.

L1-L2 and L2-L1 translation tests were scored in terms of the percent correct translations recalled for each participant, learning condition, and vocabulary type. A Vimmi word was considered correct if the two independent raters agreed that the word that was written down was valid for the sound pronounced in the audio file according to German sound-letter-mapping. A German word was considered correct if a participant wrote down the German word that was assigned to the Vimmi word during learning or if a participant wrote down a synonym of the German word, according to a standard German synonym database (http://www.duden.de).

**Table S2.** Pencil-and-paper vocabulary test scores. The maximum possible score for each condition at each time point was 68 for the free recall test and 100 for the translation tests. *M* = mean score, *SE* = standard error.

| Test type | Learning type | Vocabulary type | Time point | | | |
|---|---|---|---|---|---|---|
| | | | Day 2 $M(SE)$ | Day 3 $M(SE)$ | Day 4 $M(SE)$ | Month 5 $M(SE)$ |
| Free recall test | Gesture | Abstract | 14 (1.1) | 22 (2.1) | 27 (2.9) | 23 (3.4) |
| | | Concrete | 15 (1.1) | 23 (1.5) | 31 (2.2) | 38 (4.1) |
| | Picture | Abstract | 12 (1.4) | 26 (3.1) | 33 (3.1) | 28 (3.8) |
| | | Concrete | 11 (1.3) | 24 (2.6) | 33 (2.8) | 37 (3.7) |
| Translation tests | Gesture | Abstract | 10 (3.2) | 28 (5.5) | 44 (6.7) | 32 (5.4) |
| | | Concrete | 14 (3.5) | 38 (5.3) | 56 (5.8) | 47 (5.4) |
| | Picture | Abstract | 14 (3.0) | 38 (6.3) | 55 (6.3) | 33 (5.0) |
| | | Concrete | 16 (3.2) | 37 (5.6) | 56 (5.9) | 44 (4.9) |

**Supplementary references**